\documentclass[prl,10pt,twocolumn,showpacs]{revtex4}
\usepackage[dvips]{epsfig}
\usepackage{float}
\usepackage{amsfonts}
\usepackage{amsmath}
\usepackage{amssymb}
\usepackage{enumerate}

\newcommand{\vecb}[1]{\mathbf{#1}}

\begin{document}

\title{Optical-fiber based measurement of an ultra-small volume high-$Q$ photonic crystal microcavity}

\author{Kartik Srinivasan}
\email{kartik@caltech.edu}
\author{Paul E. Barclay}
\author{Matthew Borselli}
\author{Oskar Painter}
\affiliation{Department of Applied Physics, California Institute of Technology, Pasadena, California 91125}
\date{\today}
\begin{abstract}
A two-dimensional photonic crystal semiconductor microcavity with a quality factor $Q \sim 40,000$ and a modal volume $V_{\text{eff}} \sim 0.9$ cubic wavelengths is demonstrated.  A micron-scale optical fiber taper is used as a means to probe both the spectral and spatial properties of the cavity modes, allowing not only measurement of modal loss, but also the ability to ascertain the in-plane localization of the cavity modes.  This simultaneous demonstration of high-$Q$ and ultra-small $V_{\text{eff}}$ in an optical microcavity is of potential interest in quantum optics, nonlinear optics, and optoelectronics.  In particular, the measured $Q$ and $V_{\text{eff}}$ values could enable strong coupling to both atomic and quantum dot systems in cavity quantum electrodynamics.

\end{abstract}
\maketitle

\setcounter{page}{1}

The use of an optical microcavity to alter the interaction of light with matter has been instrumental within a wide range of fields, including cavity quantum electrodynamics (cQED) \cite{ref:Kimble2}, nonlinear optics \cite{ref:Chang}, and molecular sensing \cite{ref:Norris}.  This interaction depends strongly upon the cavity photon lifetime, measured by the quality factor $Q$, and the electromagnetic energy density within the cavity, quantified by the effective modal volume $V_{\text{eff}}$ \footnote{$V_{\text{eff}}$ is defined as the electric field energy within the cavity normalized to the peak electric field energy density \cite{ref:Foresi}}.  Planar photonic crystal (PC) microcavities have attracted significant attention \cite{ref:Painter3,ref:Foresi,ref:Ryu3,ref:Srinivasan1} in this regard due to their ability to trap light within volumes approaching the theoretical limit \cite{ref:Painter3}, the potential for on-chip integration with waveguides \cite{ref:Noda2,ref:Smith4,ref:Barclay2}, and lithographic control of many salient properties of the cavity modes \cite{ref:Painter12}.  A major limitation, however, has been the relatively low experimentally-realized $Q$ values, limited to $< 3,000$ \cite{ref:Yoshie2} until recent demonstrations of $Q \sim 6,400$ in an add-drop filter \cite{ref:Akahane} and $Q \sim 13,000$ \cite{ref:Srinivasan4} in a sub-threshold laser cavity.  In this Letter, we present optical fiber taper measurements of a PC microcavity supporting a mode with $Q \sim 40,000$ and in-plane localization consistent with $V_{\text{eff}} \sim 0.9$ cubic wavelengths ($(\lambda/n)^3$).  This simultaneous realization of a high-$Q$ {\emph{and}} ultra-small-$V_{\text{eff}}$ PC microcavity is of importance to a number of applications in the aforementioned disciplines, while the fiber taper probe is of potential value to future studies of wavelength-scale resonators.  

\begin{figure}[ht]
\begin{center}
\epsfig{figure=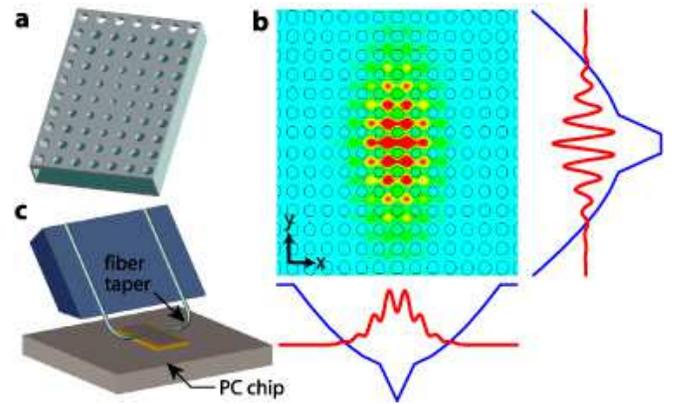, width=\linewidth}
\caption{Design of the photonic crystal (PC) membrane microcavity and experimental measurement setup. (a), Schematic of the undercut, two-dimensional PC microcavity geometry. (b), Magnetic field amplitude ($|\vecb{B}|$) in the center of the PC membrane for the  $A^{0}_{2}$ mode.  The blue curves show the grade in hole radius ($r/a$) along the central $\hat{x}$ and $\hat{y}$ axes of the cavity, and the red curves are slices of the scalar field component $B_{z}$ along these directions.  The hole radius ($r$) varies between $r/a$=0.23 in the center to $r/a$=0.35 at the edges of the cavity, and for the mode of interest, $a/\lambda_{c}$=0.245. (c), Schematic of the fiber taper probe measurement setup.}
\label{fig:cavity_design}
\end{center}
\end{figure}

The PC cavity geometry employed is shown in Figure \ref{fig:cavity_design}a,b, and was designed using group theoretical, Fourier space, and finite-difference time-domain (FDTD) analyses as described in detail elsewhere \cite{ref:Srinivasan1}.  The cavity consists of a localized defect in a square lattice of air holes that are etched into an optically thin membrane of refractive index $n\text{=}3.4$.  This geometry provides in-plane modal localization via distributed Bragg reflection due to the lattice and vertical confinement by total internal reflection at the membrane-air interface.   The resulting TE-like (electric field in the plane of the slab) $A^{0}_{2}$ defect mode (so labelled due to its symmetry and ground-state frequency in the cavity) shown in Fig. \ref{fig:cavity_design}b is predicted to have $Q \sim 10^5$ and $V_{\text{eff}} \sim 1.23$ $(\lambda_{c}/n)^3$, where $\lambda_{c}$ is the wavelength of the resonant cavity mode.  The important aspects of the cavity design are: (1) the dominant electric field component, $E_{x}$, is odd about the $\hat{x}$-axis, thereby reducing vertical radiation loss from the patterned slab, (2) a grade in the hole radius is used to both further confine the mode in-plane and reduce in-plane radiative losses, and (3) the design is relatively insensitive to perturbations to the cavity, as verified through simulations where the steepness of the grade and the average hole radius ($\bar{r}$) have been varied significantly without degrading the $Q$ below $\sim 20,000$.  

\begin{figure}[ht]
\begin{center}
\epsfig{figure=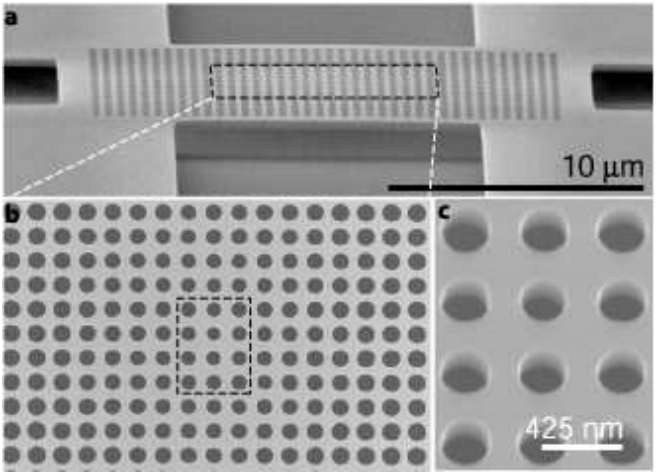, width=\linewidth}
\caption{SEM micrographs of a fully fabricated PC microcavity. (a), Cross-sectional view. (b), Top view of the portion of the cavity contained within the dashed lines in (a). Total cavity dimensions are $\sim13\mu$m x $16\mu$m.  (c), Zoomed in angled view of the dashed line region in (b) showing the smoothness and verticality of the etched air hole sidewalls, necessary to limit scattering loss and radiative coupling to TM-like modes.}
\label{fig:SEM_combo}
\end{center}
\end{figure}

In this work, PC membrane microcavities are formed from a silicon-on-insulator wafer consisting of a $340$ nm thick silicon (Si) layer on top of a 2 $\mu$m silicon dioxide layer (although Si was chosen here, similar high refractive index PC microcavities have been fabricated in a wide range of semiconductors, including AlGaAs-based \cite{ref:Yoshie2} and InP-based \cite{ref:Srinivasan4} systems).  Cavities are fabricated using electron beam lithography, a plasma etch through the Si layer, and a hydrofluoric acid wet etch to remove the underlying oxide layer.  Fully processed chips consist of a linear array of cavities, with lattice constant ($a$) varying between $380$-$430$ nm and with $5$-$10$ different lattice filling fractions (or equivalently, $\bar{r}$) for a given $a$.  Scanning electron microscope (SEM) images of a fabricated device are shown in Fig. \ref{fig:SEM_combo}.        

\begin{figure}[ht]
\begin{center}
\epsfig{figure=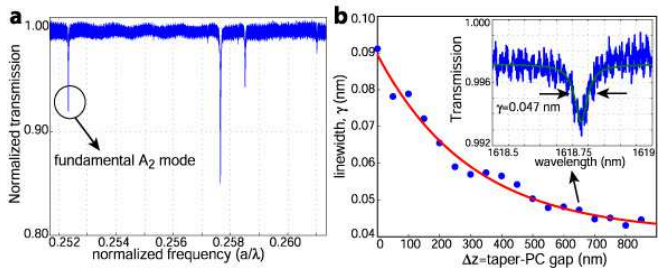, width=\linewidth}
\caption{Fiber taper transmission measurements of the PC microcavities. (a), Taper transmission spectrum for a cavity with $a$=409 nm, where the data has been normalized to a background spectrum when the taper is far above the cavity. The highlighted long wavelength mode is the $A^{0}_{2}$ resonant mode.  (b), Measured linewidth (blue dots) versus taper-cavity gap (${\Delta}z$) for the $A^{0}_{2}$ mode ($a/\lambda_{c} \sim 0.263$) in a sample with $a$=425 nm.  The taper is vertically positioned by a stepper motor with $50$ nm encoder resolution.  The red curve is a fit to the experimental data. (inset) Normalized taper transmission versus wavelength when the taper is $650$ nm above the cavity surface.}
\label{fig:z_scan_data}
\end{center}
\end{figure}

To study the resonant modes of the PC microcavities an optical fiber-based probe is used.  Optical fiber tapers \cite{ref:Knight,ref:Birks} are fabricated by heating and stretching a standard single mode fiber to a diameter of $1$-$2$ $\mu$m, providing an air-guided fundamental mode with an evanescent tail extending into the surrounding air.  The taper is mounted so that it can be laterally and vertically positioned over the cavities (Fig. \ref{fig:cavity_design}c), and is connected to a scanning tunable laser with 1 pm resolution and a $\lambda=1565$-$1625$ nm wavelength scan range.  When the taper is brought close ($\sim500$ nm) to the surface of a given cavity, a number of resonance features in the taper transmission appear (Fig. \ref{fig:z_scan_data}a).  These features disappear if the taper is positioned to the side of the cavity or if the polarization of the input light is adjusted to be TM-polarized relative to the PC slab.  The resonance positions shift as a function of $a$ and $\bar{r}$ in a manner consistent with that expected for modes supported by the photonic lattice \cite{ref:Joannopoulos}.  Noting that the defect mode of Fig. \ref{fig:cavity_design} is the lowest frequency mode lying within the (partial) photonic band-gap of the lattice, an experimental spectral identification of this mode is made for fixed $a$ as follows.  Devices are tested as a function of decreasing $\bar{r}$, so that initially, when the devices have large-sized holes, all of the resonant modes formed from the appropriate photonic band-edge lie above the frequency scan range of the input laser.  We then test in succession cavities with decreasing $\bar{r}$ until a resonance is observed.  This resonance is the lowest frequency mode lying within the in-plane (partial) bandgap of the photonic crystal, and thus corresponds to the cavity mode of interest, the \emph{fundamental} $A^{0}_{2}$ mode.

A wavelength scan of the taper transmission showing the resonance of the $A^{0}_{2}$ mode for a device with $a$=425 nm is given in the inset of Fig. \ref{fig:z_scan_data}b. In this measurement the taper is positioned parallel to the $\hat{y}$-axis at a height of $\Delta z = 650$ nm above the center of the PC microcavity.  Fitting this transmission data to a Lorentzian, a linewidth $\gamma \sim 0.047$ nm is measured for the cavity resonance.  This linewidth is a \emph{maximum} estimate for the cold-cavity linewidth $\gamma_{0}$ due to cavity loading effects of the taper.  Loading by the taper results in out-coupling to the \emph{forward} propagating fundamental taper mode which, upon interference with the power directly transmitted past the cavity, results in the observed resonant feature in the taper transmission.  Other parasitic taper loading effects include coupling to radiation modes, higher-order taper modes, and the \emph{backward} propagating fundamental taper mode.  To estimate the taper loading effects on the $A^{0}_{2}$ cavity mode, we examine $\gamma$ as a function of ${\Delta}z$.  The resulting data (Fig. \ref{fig:z_scan_data}b) shows that as ${\Delta}z$ increases, the loading effects are reduced, until a regime is reached where the taper does not significantly effect the cavity mode and the measured linewidth asymptotically approaches the cold-cavity linewidth.  Assuming that the loading is monoexponentially related to ${\Delta}z$, we fit the measured linewidth to the function $\gamma=\gamma_{0} + \beta e^{-\alpha \Delta z}$, where $\gamma_{0}$, $\beta$, and $\alpha$ are all fitting parameters.   The resulting fit value of $\gamma_{0}$ is $0.041$ nm, essentially identical to the directly measured linewidth when ${\Delta}z \gtrsim 800$ nm, and corresponds to a cold-cavity $Q \sim 39,500$.  To compare this result directly with numerical calculations, we repeat our previous FDTD calculations \cite{ref:Srinivasan1} but include an offset in $\bar{r}$ of $r/a=0.05$ to account for the increased size of the fabricated holes (as measured by SEM) relative to the design of Fig. \ref{fig:cavity_design}.  Doing so yields a predicted $Q \sim 56,000$ and $a/\lambda_{c} \sim 0.266$, fairly close to the measured values, and a $V_{\text{eff}}=0.88 (\lambda_{c}/n)^3$, smaller than the original design due to the better in-plane confinement provided by the larger hole radii.  As a final comment on $Q$, we note that a number of devices encompassing a range of values for $\bar{r}$ and the grade in $r/a$ have been tested, and $Q$ values $\gtrsim 15,000$ have been consistently measured, in accordance with simulation results.  As described earlier, this robustness to deviations from the ideal structure is perhaps the most important aspect of the cavity design.

\begin{figure}[ht]
\begin{center}
\epsfig{figure=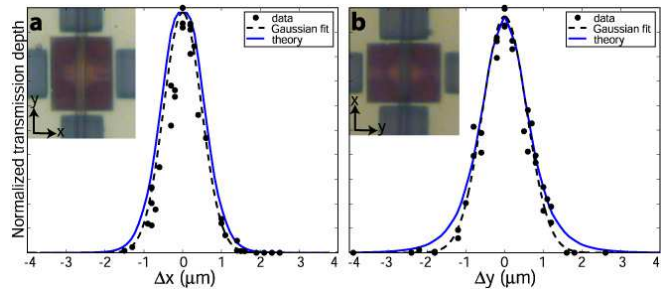, width=\linewidth}
\caption{Mode localization data. Measured normalized taper transmission depth (dots) as a function of taper displacement along the (a) $\hat{x}$-axis and (b) $\hat{y}$-axis of the cavity. The dashed line in (a)-(b) is a Gaussian fit to the data while the solid line is a numerically calculated coupling curve based upon the FDTD cavity field and analytically determined fundamental fiber taper mode (taper diameter $\sim 1.7$ $\mu$m).  The Gaussian fit to the data gives a full-width at half-maximum (FWHM) of $1.2$ $\mu$m and $1.3$ $\mu$m along the $\hat{x}$ and $\hat{y}$ axes, respectively.  The insets in (a)-(b) are optical micrographs of the taper aligned along the $\hat{y}$ and $\hat{x}$ axes of the cavity, respectively.  The cavity is the central reddish-brown rectangular region.}
\label{fig:lateral_scan_data}
\end{center}
\end{figure}

By measuring (for fixed ${\Delta}z$) the strength of the coupling to the PC cavity modes as a function of lateral taper displacement, the in-plane localization of the cavity modes can be ascertained \cite{ref:Knight2}, \footnote{Here the same optical fiber taper near-field probe is used to both excite the PCWG modes and to map their spatial profile.  Other works employing evanescent coupling from eroded monomode fibers to excite silica microsphere whispering-gallery modes have used a secondary fiber tip to collect and map the mode profiles \cite{ref:Knight2}.}.  The strength of coupling is reflected in the \emph{depth} of the resonant dip in the taper transmission \footnote{The maximum transmission depth achieved for the mode of interest was $\sim10\%$, though coupling to other modes reached depths as large as $\sim30\%$.  Coupling in all cases was limited to the \emph{under-coupled} regime \cite{ref:Cai}.}.  For the taper aligned along the long ($\hat{y}$) and short ($\hat{x}$) axes of the cavity the depth of the resonant transmission dip for the $A^{0}_{2}$ cavity mode versus taper misalignment is shown in Figs. \ref{fig:lateral_scan_data}a and \ref{fig:lateral_scan_data}b, respectively.  These measurements show the mode to be well-localized to a micron-scale central region of the cavity, and confirm that the $A^{0}_{2}$ mode is both high-$Q$ and small $V_{\text{eff}}$.  They do not, however, reveal the highly oscillatory cavity near-field, but rather an envelope of the field, due to the relatively broad taper field profile.  To better understand the results of Fig. \ref{fig:lateral_scan_data}, we consider a simple picture of the taper-PC cavity coupling \cite{ref:Haus_book}, where the coupling coefficient is calculated from the analytically-determined taper field and the phase-matched Fourier components of the FDTD-generated cavity field.  The calculated resonant transmission depth as a function of taper displacement is shown in Figs. \ref{fig:lateral_scan_data}a,b as solid lines and agrees closely with the measured data.  Assuming that the cavity mode is localized to the slab in the $\hat{z}$-direction, the close correspondence between the measured and calculated in-plane localization indicates that $V_{\text{eff}} \lesssim 0.9 (\lambda_{c}/n)^3$ for this high-$Q$ mode.  Similar measurements of the higher-frequency resonant modes of the PC microcavity (such as those in Fig. \ref{fig:z_scan_data}a) indicate that they are significantly more delocalized in-plane in comparison to the $A^{0}_{2}$ mode, as one might expect for higher-order modes of the cavity.       

To illustrate the potential applications of such a small $V_{\text{eff}}$ and high-$Q$ microcavity, we consider two examples from quantum optics.  The Purcell factor ($F_{P}$), a measure of the microcavity-enhanced spontaneous emission rate of an embedded active material, is given under suitable (maximal) conditions as \cite{ref:Gayral}:         

\begin{equation}
\label{eq:Purcell}
F_{P}=\frac{3}{4\pi^2}\Bigl(\frac{\lambda_{c}}{n}\Bigr)^3\Bigl(\frac{Q}{V_{\text{eff}}}\Bigr).    
\end{equation}

\noindent For the PC microcavity studied here ($Q \sim 40,000$, $V_{\text{eff}} \sim 0.9 (\lambda_{c}/n)^3$), the predicted $F_{P}$ is $\sim 3,500$, an extremely large value for a semiconductor-based microcavity (previous work on semiconductor microdisks \cite{ref:Gayral} have predicted $F_{P}\sim190$).  Another application is in cQED, where strongly coupled atom-photon systems have been proposed as candidates to produce the quantum states required for quantum computing applications \cite{ref:Mabuchi}.  For such applications, the regime of strong coupling \cite{ref:Kimble2}, where the atom-photon coupling coefficient ($g$) exceeds the cavity and atomic decay rates ($\kappa$ and $\gamma_{\perp}$, respectively), must be reached.  Although strong coupling has been achieved in systems consisting of an alkali atom and an actively-stabilized Fabry-Perot cavity \cite{ref:Kimble2}, in future applications, where higher levels of integration are sought, chip-based cavities are of interest \cite{ref:Mabuchi}.  Using the measured $Q$ and estimated $V_{\text{eff}}$ for the $A^{0}_{2}$ mode studied here, the relevant parameters for a commonly-used Cesium (Cs) atomic transition ($\lambda_{0}=852$ nm, $\gamma_{\perp}=2.6$ MHz) \cite{ref:Kimble2}, and the formulas $g=\eta(\gamma_{\perp}(c\lambda_{0}^2/8\pi\gamma_{\perp}V_{\text{eff}})^{1/2})$ and $\kappa=\omega/4{\pi}Q$, we calculate $g\sim 17$ GHz \footnote{As $V_{\text{eff}}$ is defined relative to peak electric field energy density, rather than electric field strength, a factor $\eta$ must be included for dielectric cavities where the two values are not equal.  $\eta \sim 0.42$ for our cavity.} and $\kappa \sim 4.4$ GHz, indicating that the coupled Cs-PC cavity system could achieve strong coupling.  In addition, the calculated critical atom number ($N_{0}=2\kappa\gamma_{\perp}/g^2$) and saturation photon number ($m_{0}=\gamma_{\perp}^2/2g^2$) are $N_{0}\sim8.4\text{x}10^{-5}$ and $m_{0}\sim1.2\text{x}10^{-8}$, a regime where a single atom would have a profound effect on the cavity field, and vice versa.  A similar calculation for an InAs semiconductor quantum dot \cite{ref:Becher1} indicates that the current PC microcavity would also be capable of reaching strong coupling in such a solid-state system.

In conclusion, while silica fiber tapers have been successfully used to probe \emph{larger} ($V_{\text{eff}} \gtrsim 500 (\lambda_{c}/n)^3$) \emph{silica-based} resonators, such as microspheres \cite{ref:Knight,ref:Cai} and microtoroids \cite{ref:Armani}, the current work illustrates the use of tapers as a probe for \emph{ultra-small mode volume{\text{,}} high refractive index} ($n\sim3.4$) cavities, where the micron-scale dimension of the taper is utilized to both source and out-couple light from the cavity.  This technique allows for rapid characterization of relevant cavity mode parameters ($\lambda_{c}$, $Q$, and $V_{\text{eff}}$), and although it does not necessarily provide efficient coupling to the cavity [28], as a suitably designed waveguide may \cite{ref:Barclay2}, the use of an external fiber-based probe provides a greater level of versatility than other methods, such as embedding of active material within the cavity \cite{ref:Yoshie2,ref:Srinivasan4} or microfabrication of input-output waveguides to couple to the cavity \cite{ref:Noda2,ref:Lin4}.  In particular, resonant cavity elements in both passive and active devices can be tested, making it a valuable tool for future studies of wavelength-scale resonators.  Here, we have used this technique to demonstrate a PC microcavity mode with $Q \sim 40,000$ and in-plane localization consistent with $V_{\text{eff}} \lesssim 0.9$ $(\lambda_{c}/n)^3$. 

The authors thank B. Lev and H. Mabuchi for useful discussions pertaining to cQED. K.S. thanks the Hertz Foundation and M.B. thanks the Moore Foundation, NPSC, and HRL Laboratories for their graduate fellowship support.

\bibliography{./PBG}
\end{document}